\documentclass[useAMS]{mn2e}
\usepackage{graphicx}
\def\ltsima{$\; \buildrel < \over \sim \;$}
\def\gtsima{$\; \buildrel > \over \sim \;$}
\def\lsim{\lower.5ex\hbox{\ltsima}}
\def\gsim{\lower.5ex\hbox{\gtsima}}

\title[Tidal distortions around the old open cluster NGC~6791]
       {Evidence of tidal distortions and mass loss from the old open cluster NGC~6791}
\author[Dalessandro et al.]
  {E.~Dalessandro$^1$, P.~Miocchi $^1$, G. Carraro $^2$,  L. J\'{i}lkov\'{a}$^3$, A. Moitinho $^4$ \\
    $^1$Dipartimento di Astronomia, Universit\`a degli Studi
     di Bologna, via Ranzani 1, I--40127 Bologna, Italy\\
    $^2$ESO,Alonso de Cordova 3107, 19001 Santiago de Chile, Chile\\
    $^3$Leiden Observatory, P.O. Box 9513, NL-2300 RA  Leiden, The Netherlands\\
    $^4$ SIM/CENTRA, Faculdade de Ciencias de Universidade de Lisboa, Ed. C8, Campo Grande, 1749-016
    Lisboa, Portugal\\  
}

\date{23 February, 2015}

\pagerange{\pageref{firstpage}--\pageref{lastpage}} \pubyear{2014}

\def\LaTeX{L\kern-.36em\raise.3ex\hbox{a}\kern-.15em
    T\kern-.1667em\lower.7ex\hbox{E}\kern-.125emX}

\begin{document} 

\label{firstpage}

\maketitle
\begin{abstract}

We present the first evidence of clear signatures of tidal distortions in the density distribution of the fascinating open cluster 
NGC~6791. We used deep and wide-field data obtained with the Canada-France-Hawaii-Telescope covering a $2^{\circ} \times 2^{\circ}$ 
area around the cluster.
The two-dimensional density map obtained with the optimal matched filter technique shows a clear elongation and 
an irregular distribution starting from $\sim300\arcsec$ from the cluster center. 
At larger distances, two tails extending in opposite directions beyond the tidal radius are also visible.
These features are aligned to both the absolute proper motion and to the Galactic center directions.
Moreover, other overdensities appear to be stretched in a direction perpendicular to the Galactic plane.

Accordingly to the behaviour observed in the density map, we find that 
both the surface brightness and the star count density profiles reveal a departure from a King model 
starting from $\sim600\arcsec$ from the center. 

These observational evidence suggest that NGC~6791 is currently experiencing mass loss likely due to 
gravitational shocking and interactions with the tidal field. We use this evidence to argue that NGC~6791 should 
have lost a significant fraction of its original mass.   
A larger initial mass would in fact explain why the cluster survived so long.
Using available recipes based on analytic studies and $N$-body simulations, we derived the expected mass loss due to stellar
evolution and tidal interactions and 
estimated the initial cluster mass to be $M_{\rm ini}=(1.5$--$4) \times 10^5\,\mathrm{M}_{\odot}$.
\end{abstract} 

\begin{keywords}
Open clusters: individual (NGC~6791); stars: evolution; stars: imaging
\end{keywords}

\section{Introduction}		

NGC~6791 is one of the most intriguing open cluster (OC) in the Galaxy. It possesses a unique set of
properties that have attracted a lot of attention in the last decade. 

Indeed, NGC~6791 is one of the most massive ($M\sim 5000 M_{\odot}$; Platais et al 2011), oldest ($t\sim8$ Gyr; King et al. 2005; 
Grundahl et al. 2008; Garcia-Berro et al. 2010; Brogaard et al. 2012) and most metal-rich ([Fe/H] $\sim$ +0.40; 
Carraro et al. 2006, Gratton et al. 2006, Origlia et al. 2006) Galactic OCs. More specifically, this combination of 
properties is unique for clusters at its Galactocentric distance ($d\sim 8 Kpc$; Carraro 2014).
Stars with similar age and metallicity are found 
in the Galactic bulge (Bensby et al. 2013) where it has been suggested (J\'{i}lkov\'{a} et al. 2012)  NGC~6791 might have formed.

Given its peculiarities, and in particular its mass, NGC~6791 represents an ideal link between OCs and globular clusters. 
For this reason it has been widely targeted to understand whether 
star-to-star variations of light element (C, N, O, Na, O, Mg and Al) abundances can be found also in this system as observed 
in basically all globular clusters (Gratton et al. 2012).  
This topic is still matter of debate and no consensus has been reached so far on whether the cluster hosts more 
than one 
generation of stars. Geisler et al. (2012) presented evidence of Na-O anti-correlation among giant stars in NGC~6791,
while recently Bragaglia et al. (2014) 
and Cunha et al. (2015) have questioned this result finding an homogeneous Na abundance (within $\sim 0.1$ dex).

NGC~6791 shows also an anomalous horizontal branch (HB) with a well populated red clump, 
as expected for its high metallicity, and
a group of hot HB stars (Liebert et al. 1994, Tofflemire et al. 2014), whose origin is still not completely understood (Carraro 2014). 
The combined UV flux of the few hot HB stars and its metallicity makes this cluster a fairly good proxy 
of standard UV-upturn elliptical galaxies (Buzzoni et al. 2012).  
This result is supported also by the Lick indices analysis of its spectral energy distribution.

This plethora of unique properties makes NGC 6791 an extremely interesting object to study and understand.
A number of questions about the nature of this system still arise:  
how and where could such a stellar system have formed? Is NGC 6791 a genuine OC? Did it form close to the bulge?
How could have survived for so long in the adverse high-density environment of the inner Galactic disk, where it presently is?
One of the still missing key ingredient that could help to understand the origin of such a fascinating system 
is its original mass. 

The mass budget of any stellar system is strongly affected by two-body relaxation, stellar evolution, encounters with 
giant molecular clouds and spiral arms, interactions with the tidal field.
The disruption scenarios and the survival rates of Galactic OCs have been studied since the sixties (Spitzer \& Harm 1958;
Spitzer 1958).
Recently, $N$-body simulations (e.g., Gieles et al. 2006; Gieles \& Baumgardt 2008) have also shown the impact of giant molecular 
clouds and of the initial structure of OCs on their dissolution. 
The OC orbits have generally small eccentricities and they tend to be located at a small height ($Z$) on the Galactic plane,
as a consequence they are expected to pass many times trough the Galactic disk. Each of these gravitational shocks heat up 
and compress the cluster 
which then takes an elongate shape which flattens to its maximum at $Z=0$ (Leon 1998). 
Repeated disk-shocking speed-up the disruption of 
OCs (Combes et al. 1999) and  after each passage cluster stars rejected in the halo of the system are stripped out by 
the gravitational field of the Galaxy. These ejected stars are expected to form tidal tails extending far from the cluster inner regions. 
Indeed, Davenport \& Sandquist (2010) and Bergond et al. (2001) have found that some OCs 
show a quite elongated and extended structures likely resulting from the interactions with the Galactic tidal field. 

In general, OCs are thought to dissipate in a timescale of the order of $\sim 10^7 - 10^8$ yr (Binney \& Tremaine 1987).
In this context the existence of very old OCs like NGC~6791 represents an anomaly\footnote{
Other old ($t>2-3$ Gyr) OCs for which such an argument applies are NGC~188, M~67 and Berkley~11.}. 
This system may have managed to survive so long either because it moved along a preferential orbit or 
its original mass was much larger than observed today.

In this paper we analyse the two-dimensional (2D) and projected density distributions of NGC~6791 with the aim 
of understanding 
whether the cluster suffered from strong tidal interactions that could have
significantly decreased its mass. 
We find that this system has an elongated structure and shows overdensities in the external regions
that are aligned to the direction of the orbital motion. These features 
strongly suggest 
that this system has likely lost part of its mass because of gravitational shocks and interaction with the Galactic 
tidal field.
We use this argument to support the hypothesis that this system was significantly more massive than observed today and we derive
an estimate for its original mass by means of a simplified analytic approach.

The paper is organised as follows: 
in Section~2 data-set and data-reduction procedures are described, in Section~3 the 2D density map is obtained  while in
Section~4 both the star-count projected density and the surface brightness profiles are analysed; finally in Section~5 
we estimate, the initial mass of NGC~6791.

\section{Observations and data reduction.}

\begin{figure}
\includegraphics[width=85mm]{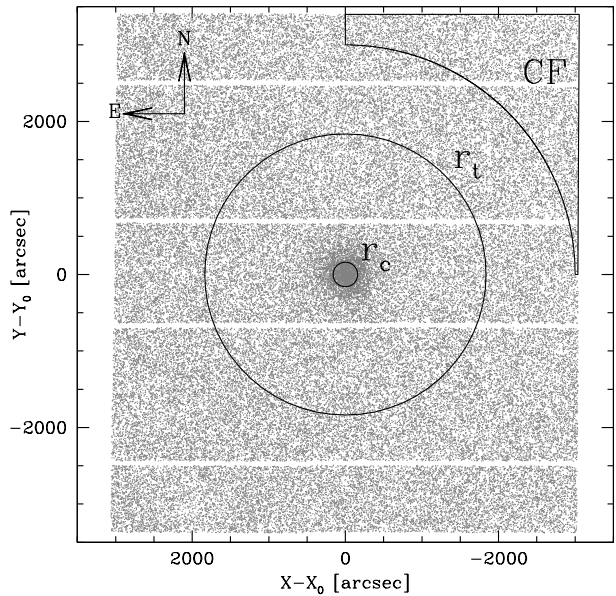}
\caption{Map of the entire CFHT database with respect to the position of 
C$_{\rm grav}$ (X$_0$,Y$_0$). The two circles represent the location of core ($r_c$) and tidal radii ($r_t$). The box in
the upper right corner defines the "control field" (CF) area.}
\label{map}
\end{figure}

We used deep proprietary images (Prop ID: OPTICON 2013B/005; PI Moitinho) obtained 
with the wide field imager MegaCam mounted at the Canada-France-Hawaii Telescope (CFHT).
This camera consists of a mosaic of 36 chips of $2048\times 4612$ pixels each, 
with a pixel scale of $0.185\arcsec$ pixel$^{-1}$ providing a total field of 
view of $\sim 1^{\circ}\times1^{\circ}$.
We used four different pointings set up to cover a total field of view of $\sim
2^{\circ}\times2^{\circ}$
centered approximately on the cluster center (Figure~1). 
For each pointing four images acquired both with the $g'$ and $r'$ bands and 
with exposure times 
$t_{\rm exp}=100$ sec and $t_{\rm exp}=180$ sec respectively have been secured. 
An appropriate dither pattern of few arcesc have been adopted for each pointing. This allowed us to cover 
most of the inter-chip gaps, with the exception of the most prominent and horizontal ones (see Figure~1).

The data were pre-processed (i.e. bias and flat-field corrected) by the Elixir pipeline developed 
by the CFHT team.
By means of an iterative procedure, an adequate number ($>20$) of isolated and bright stars have been 
selected in each chip and band to model the Point Spread Function (PSF). Then the PSF model 
was applied to all the stellar-like sources at about $4\sigma$ from the local background by using 
DAOPHOT and the PSF-fitting algorithm ALLSTAR (Stetson 1987). 
For each filter and chip we matched the single-frame catalogues to obtain a master list. 
Finally, the $g'$ and $r'$ master lists were then geometrically matched and combined.
Each  master list includes the instrumental magnitude, defined as the weighted mean of the single image
measurement reported to the system reference frame of the transformation, and the error, which is the standard
error of the mean. 

The instrumental coordinates have been reported 
independently for each chip to the absolute astrometric system  ($\alpha$, $\delta$) using a large 
number of stars in common
with the DR7 release of the Sloan Digital Sky Survey (Abazajian et al. 2007) and following the procedure 
described by Dalessandro et al. (2009). Using the same stars, we derived for each chip and band the zero-points 
to report the instrumental magnitudes to the Sloan photometric system. At this stage all the master lists are in the same photometric
and astrometric system. We then merged them to create the final catalogue, which counts about 270,000 stars.
 
The ($g'$, $g'-r'$) colour-magnitude diagrams (CMDs) of NGC~6791 and of the surrounding field are
shown in Figure~2.
In the CMDs, it is possible to clearly distinguish the Main Sequence (MS) of the cluster 
(left panel of Figure~2), 
extending down to $g'\sim23.5$ in the innermost regions, corresponding to about five magnitudes below the Turn-Off.
Also, it is possible to note that our data suffer of strong saturation starting at $g'\sim17$.
Two almost vertical sequences located around $(g'-r')\sim0.6$ and $(g'-r')\sim1.5$ are clearly visible 
both in the innermost (Figure~2 left panel) and external regions (Figure~2 right panel). 
They are populated by M-dwarf and Main-Sequence Galactic disk stars.

A differential reddening of about $\Delta E(B-V)\sim0.05$ has been estimated for this cluster 
(see for example Platais et al. 2011). Indeed, at the level of the Turn-Off ($g'\sim18$), the MS shows a mild broadening 
not explainable in terms of pure photometric errors. Such an effect disappears at fainter 
magnitudes ($g'\sim19$), where the MS starts to bend diagonally in the ($g'$, $g'-r'$) CMD. 
Given its relatively small amplitude, we will neglect the effect of differential reddening in the following analysis.

\section{The density map}
\subsection{The matched filter analysis}

\begin{figure}
\includegraphics[width=85mm]{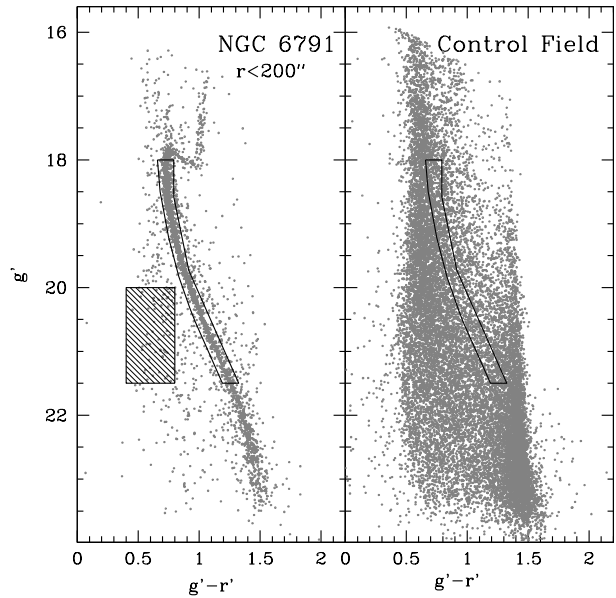}
\caption{($g'$, $g'-r'$) CMDs of the innermost region of NGC~6791 ({\it left panel}) and 
of the "Control Field" ({\it right panel}). The black box along the MS of NGC~6791 encloses the region 
in the CMD where CP stars have been selected. The shaded area defines the region where fiducial Galactic disk MS stars
have been selected.}
\label{map}
\end{figure}

To probe the spatial distribution of NGC~6791 we first
analysed the 2D density distribution.

The density analysis of NGC~6791 has been performed in the circular area enclosed within a distance from the cluster center (see
details in Section~4) $r=3000\arcsec$,
thus covering a $1.7^{\circ}\times1.7^{\circ}$ area. 
We selected stars along the MS in the magnitude range $18<g'<21.5$ and
within three times the local color dispersion about the cluster mean ridge line (see Figure~2). 
Stars selected in this way define the ''Cluster Population" (CP) sample.
These selection criteria have been chosen
to ensure a good level of completeness and to minimize the contamination from Galactic field stars. 
Variations of the adopted magnitude limits do no affect the final result.

As a first qualitative step, we obtained a rough smoothed density distribution of CP stars.
As expected, at a first look, the distribution of CP stars appears to be extremely concentrated 
around the position of the cluster and it shows some irregularities and structures allover the field of view (FOV). 
We also note that the North-West quadrant shows on average a smoother distribution than the other quadrants for $r>1000\arcsec$.

To investigate in more detail the two-dimensional distribution of CP stars, 
we adopted the widely used optimal matched filter
technique (see for example Kuhn et al. 1996; Rockosi et al. 2002; Odenkirchen et al. 2001, 2003).
It has been shown (Davenport \& Sandquist 2010) that this method 
is robust for identifying low-density stellar populations against a significant background, as in the case of
OCs.
The optimal matched filter technique requires to use a fiducial field population as reference. 
For this purpose, we selected stars in the 
same magnitude and colour bins as the CP sample and located 
at $r>3000\arcsec$ from the
cluster center in the North-West quadrant (because of its apparent homogeneous distribution). We named this sample of stars 
``Control Field" (CF; Figures~1 and 2 ).

We computed the densities in the CMDs (Hess diagrams) of both the CP and CF samples.
Then we assigned to each star in the CP sample
a weight defined as the reciprocal of the fraction of CF stars lying within a fixed range of colour and magnitude  
from its location in the CMD.

\begin{figure*}
\includegraphics[width=110mm]{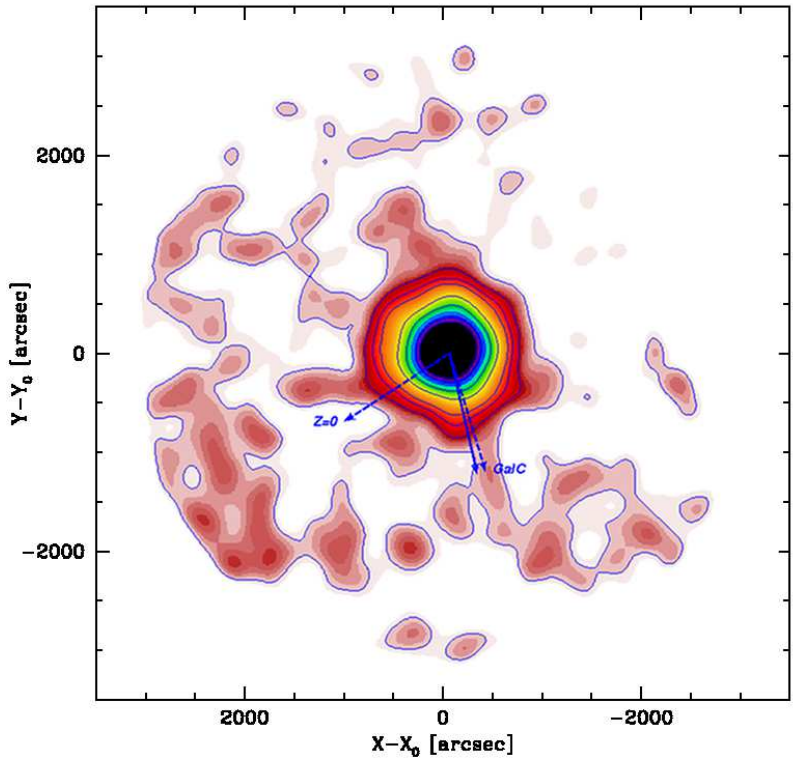}
\caption{Large-scale two-dimensional colour-coded surface density map around NGC~6791. 
The contour levels span from $3\sigma$ to $40\sigma$ with irregular steps. The solid arrow represents the
direction of the absolute proper motion of NGC~6791, while the dashed ones mark the direction of the
Galactic center (GalC) and that perpendicular to the Galactic plane (Z=0). }
\label{map}
\end{figure*}

Finally, the weighted distribution of star positions was transformed into a smoothed surface 
density function through the use of a kernel
whose width has been chosen to be large enough so as to make negligible the impact of the gaps in the FOV.
This procedure yields the surface density distribution shown in Figure~3. 
As apparent, the overall density distribution is elongated and irregular. 
In particular, we observe that the innermost $\sim 300\arcsec$
are quite spherical, while at larger distance the cluster starts to be distorted mainly along the 
North-East/South-West direction in an almost symmetrical way respect to the center of the cluster.
At $r>1000\arcsec$, two tails running at opposite directions are clearly visible, with the one 
in the southern direction
extending up to distances larger than $2500\arcsec$ from the cluster center.
It is important to note that both the external tidal arms and the innermost elongation of the cluster 
run almost parallel 
to the absolute proper motion vector (Bedin et al. 2006) and to the direction of the Galactic 
center (GalC; Figure~3). 
This behaviour is in agreement with a real tidal nature
for the observed distortions. In fact, the orbital motion of NGC~6791 likely develops close to the
Galactic equatorial plane (J{\'{\i}}lkov{\'a} et al. 2012) and, since tidal tails are 
supposed to lie on the orbital plane (at least in axisymmetric potentials, see Montuori et al. 2007),
they are expected to be seen with such a projected alignment.
Moreover, other distortions extending almost perpendicularly to the PM direction are clearly visible and their
formation could be related to
the tidal effect due to the Galactic disk (e.g., Bergond et al. 2001).

The features observed in the 2D density distribution of NGC~6791 
are indeed typical evidence of recent stellar mass loss due to dynamical evolution of the system under 
the tidal influence of the overall Galactic potential.
It is interesting to note that the density map of NGC~6791 is qualitatively similar to that
observed in M~67 by Davenport \& Sandquist (2010) and to a weaker extent to other three
OCs (namely NGC~2287, NGC~2516 and NGC~2548) by Bergond et al. (2001).

We also note that in the South-East quadrant for $r>2000\arcsec$  there is a star density increase 
that does not seem to be directly connected to the cluster. Interestingly enough, 
this overdensity appears at decreasing Galactic 
latitude therefore we speculate that it might be due to the density gradient of Galactic disk stars.

\subsection{Sanity checks}

To further check the significance of the anisotropic distribution of NGC~6791 
we used for comparison stars 
selected in the CMD in the range $0.4<(g'-r')<0.8$ and $20<g'<21.5$ (see Figure~2). These are likely MS Galactic disk 
stars and they trace the behaviour of the main contributors to the contamination of CP stars. 
This sample is supposed to be
homogeneously distributed across the FOV and actual deviations from a smooth distribution 
reflect variations in the detection efficiency, Galactic field and/or extinction gradients.

We performed on this sub-sample the same analysis done previously for CP stars.
The 2D density map of MS Galactic disk stars is shown in Figure~4. We observe that
in general they are uniformly distributed across the FOV as expected, but they show a density excess
in the external regions of the South-East quadrant. This evidence further suggests that the relevant 
feature in the 2D density distribution of CP stars at $r>2000\arcsec$ (Figure~3) in this quadrant,
 is likely due to a gradient of the density of the 
Galactic disk within the FOV.

In addition, we divided the FOV in ten angular sectors of $36^{\circ}$ each with vertex at the cluster center. 
Then 
for each sector with angle $\theta$ measured counter-clockwise from the West direction, 
and for different distances from the cluster
center we estimated the ratio between CP stars ($N_{\rm CP}$),
and likely Galactic disk MS stars, selected as described before ($N_{\rm field}$). The results for $r<500\arcsec$ and 
$500\arcsec<r<1500\arcsec$ 
are shown  in Figure~5. In both cases the ratio shows significant variations as a function of $\theta$, 
as expected for a real anisotropic density distribution. 
Moreover it is also worth noting that
different patterns in the central and outermost regions appear. 
For $500\arcsec<r<1500\arcsec$ 
the maxima of the ratio occur at $\theta=100^{\circ}$ and $\theta=280^{\circ}$ corresponding to the two
opposite main tidal arms.

\begin{figure}
\includegraphics[width=85mm]{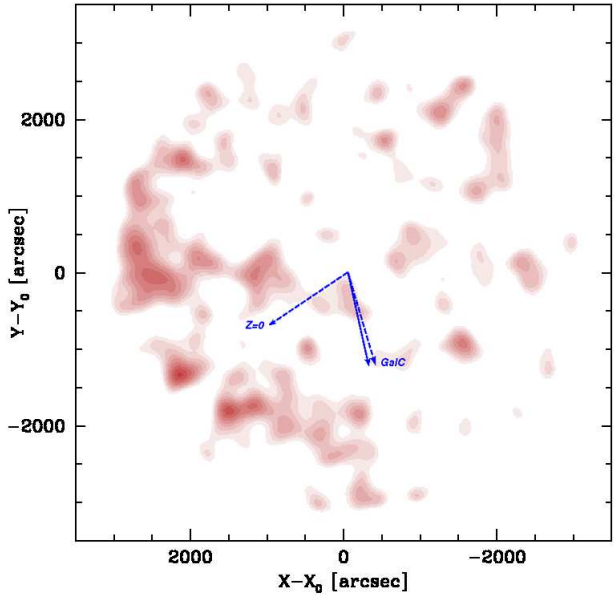}
\caption{As in Figure~3, but for stars with $0.4<(g'-r')<0.8$ and $20<g'<21.5$, which are representative of the
Galactic field contamination (see Section~3.2).
Colors are in the same relative scale as in Figure~3.}
\label{map}
\end{figure}

These sanity checks demonstrate that the observed 2D density distribution shown in Figure~3 
is a real feature of NGC~6791 and 
cannot be ascribed to observational bias or field gradients.

\section{Star density and surface brightness profiles.}

Star density distortions, elongation and tidal tails are expected to be detected also in the radial
density profile (see for example Odenkirchen et al. 2003; Capuzzo-Dolcetta et al. 2005). Therefore as a second step of our analysis, 
we studied the projected density profile of NGC~6791.

First we derived the center of gravity (C$_{\rm grav}$) by using an iterative procedure 
(see for example Dalessandro et al. 2013)
and averaging the positions $\alpha$ and $\delta$ of CP stars. 
We used the center reported by the WEBDA web page\footnote{http://www.univie.ac.at/webda/cgi-bin/ocl\textunderscore page.cgi?dirname=ngc6791}
as starting guess point of our procedure. 
To avoid spurious effects due to incompleteness we performed several estimates by adopting different
magnitude and distance selections. In particular, we used three different magnitude bins in the range 
$18<g'<21.5$, with the upper limit (fainter magnitude) increasing by 0.5 mag at each step. 
For each magnitude bin, we repeated the
procedure for four different radial selections, from $250\arcsec$ to $400\arcsec$
with a step of $50\arcsec$. We ended up with a total of 20 estimates. 
The resulting C$_{\rm grav}$ is the average of these measures and it is located at $\alpha_c=19^h 20^m 54^s.273$ and
$\delta_c = 37^{\circ} 46' 25\arcsec .20$ (R.A.= 290.2261395  
Dec=37.77366712), with an uncertainty of $\sim 1.2\arcsec$.
This new determination is located at $\sim12\arcsec$ North-East from the center reported by the WEBDA web page.
While it is not trivial to understand the origin of this discrepancy, we stress here that it has a negligible
impact on the results presented in this paper.

\begin{figure}
\includegraphics[width=85mm]{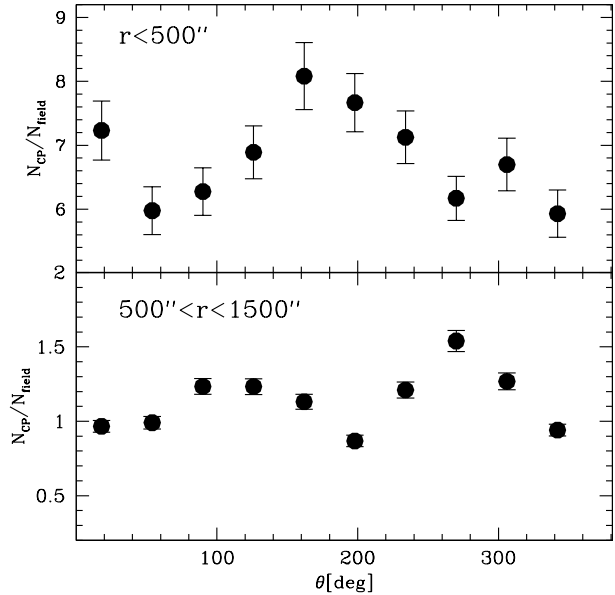}
\caption{Ratio between CP and field stars as a function of the position angle $\theta$ (Section~3) 
for stars at $r<500\arcsec$ ({\it upper panel}) and at $500\arcsec<r<1500\arcsec$ ({\it lower panel}).}
\label{map}
\end{figure}

The projected density profile of NGC~6791 has been determined using direct counts of CP stars.
We excluded the South-East quadrant from the analysis since in the most external part of this region the star distribution 
is dominated by MS Galactic disk stars (Section~3.1).
Using the procedure described in Dalessandro et al. (2013), we divided the considered 
FOV into 18 concentric annuli, each centered on C$_{\rm grav}$ and suitably split in an adequate
number of subsectors.
Number counts have been calculated in each subsector and the corresponding densities were 
obtained dividing 
them by the sampled area. Incomplete area coverage, mainly due to the presence of gaps (Figure~1), 
have been properly taken into account. 
The stellar density of each annulus was then defined as the average of the subsector densities
and its standard deviation computed from the variance among the subsectors. 
The resulting projected surface 
density profile is shown in Figure~6. 

We estimated the background density contribution by using only stars in the North-West quadrant.
In fact, as discussed before (see also Figure~3) this region is free by any evident structure caused either by 
the cluster elongation or from field stars.   
The background contribution has been estimated as the average of the four 
outermost density measurements 
($r>1400\arcsec$), which define a kind of plateau at $\log(\rho_{\rm bck})\sim -3.6$ stars arcsec$^{-2}$.
We then subtracted this value to the observed density profile.
We have verified that this choice is appropriate. In fact, 
the contribution to the background due to field stars (see selection criteria in
Section 3.1) is constant in all quadrants but the South-East one.

The derived density profile has been reproduced by using a single-mass King model (King 1966). The fitting 
procedure is fully described in Miocchi et al. (2013). 
The fit was limited to the innermost $r<500\arcsec$ to avoid the most elongated regions of the cluster.
The best-fit is obtained for a concentration $c=1.06$ (corresponding to a central dimensionless potential
$W_0=5.3$) and core radius 
$r_c=160\arcsec$.  These parameters are in partial disagreement with those obtained by Platais et al. (2011; 
$c=0.74$, $r_c=196\arcsec$) which to our knowledge is the only other density profile analysis present 
in the literature for NGC~6791.
The authors obtained the density profile using stars with $g'<22$ and weighted according to their proper motion.
However their data were limited to the innermost $900\arcsec$ and therefore the determination of the tidal radius and 
any residual in the background contribution was rather unstable.

As apparent in Figure~6, for $\log(r)>2.8$ ($r>600\arcsec$) the density profile clearly deviates 
from the behaviour predicted by the King model. The use of a
Wilson model (Wilson 1975) 
does not provide any significant improvement on the overall quality of the fit.   
We have also checked that even if we do not constrain the model to fit the innermost region, 
there is no way  to satisfactory reproduce the density profile for $r>600\arcsec$. 
The external part of the projected density
profile instead follows a power-law behaviour with an exponent $\alpha\sim-1.7$,
in agreement with the values found in Galactic globular clusters with observed tidal tails 
(see for example Sollima et al. 2011 and references therein) and as predicted by theoretical 
models (Johnston et al. 1999).
K{\"u}pper et al. (2010) analysed the variation of the 
power-law slopes of tidal tails  
of dissolving clusters as a function of the orbital phase, eccentricity  and initial 
cluster densities by means of N-body simulations. 
They found that the power-law slopes have typical values $\alpha=-4 -- -5$ that can 
increase reaching 
values $\alpha\sim-2$ for systems 
with highly eccentric orbits and close to or at the apogalacticon, 
where tidal tails get strongly compressed.  
This upper limit is in good agreement with what found for NGC~6791, however we should note that 
the analysis by K{\"u}pper et al. (2010) is performed on clusters moving on orbits perpendicular 
to the line of sight, while this is not the case for NGC~6791. As a consequence the relatively 
shallow power-law observed in NGC~6791 can be the result also of projection effects.

\begin{figure}
\includegraphics[width=85mm]{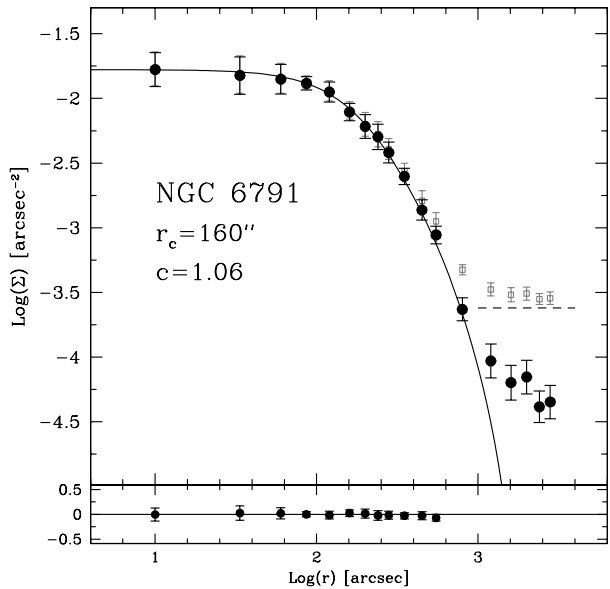}
\caption{Observed star count density profile as a function of radius (open grey squares). The dashed
line represents the density value of the background as obtained by averaging the four outermost density
measurements in the North-West quadrant. The black filled dots are densities obtained after background
subtraction. The best single-mass King model is also overplotted to the observations (solid line) and
the structural parameters are labeled. The lower panel shows the residual between the observations and
the best-fit model.}
\label{map}
\end{figure} 

To further investigate the results just described and check whether 
they might be caused by incompleteness or bias of our photometric catalogue, 
we also derived the surface brightness profile by performing the analysis directly 
on FITS images (see for example Dalessandro et al. 2012).
We used a low-resolution single-plate ESO Digitalized Sky Survey image obtained in optical band and extending up to 
about $2000\arcsec$ from C$_{\rm grav}$. We limited the analysis to the same quadrants 
used for deriving the star counts density profile and we used the North-West quadrant 
to estimate the background contribution.
We divided the FOV in a similar number of concentric annuli and subsectors. The surface brightness 
assigned to  each annulus is given by the ratio between the mean of the fluxes measured in the 
subsectors and the area covered by the annulus. 
The resulting surface brightness profile is shown in Figure~7. First we observe that 
the King model best-fitting the star counts density profile well reproduces also the 
surface brightness profile.
More importantly, we note that consistently with what found in the star count projected density profile, 
for $r>600\arcsec$ the surface brightness profile 
starts to deviate from the behaviour predicted by the model and it declines as a power-law. 

We can therefore safely conclude that both the 2D density map and the density 
(surface brightness) profile analysis
suggest that NGC~6791 is currently experiencing tidal stripping events.

It is worth noting that King et al. (2005) observed a rather flat mass function 
 in the centre of NGC~6791 by using very deep {\it Hubble Space Telescope} observations.
 This observational evidence further supports the results obtained in this work.
 In fact, for clusters that experienced episodes of mass loss, a flat mass function is 
 expected because of the preferential depletion of low mass stars.   
 A detailed analysis of the radial variation of the mass function of this system 
 will be presented in an upcoming paper (E. Dalessandro et al. 2015).

\begin{figure}
\includegraphics[width=85mm]{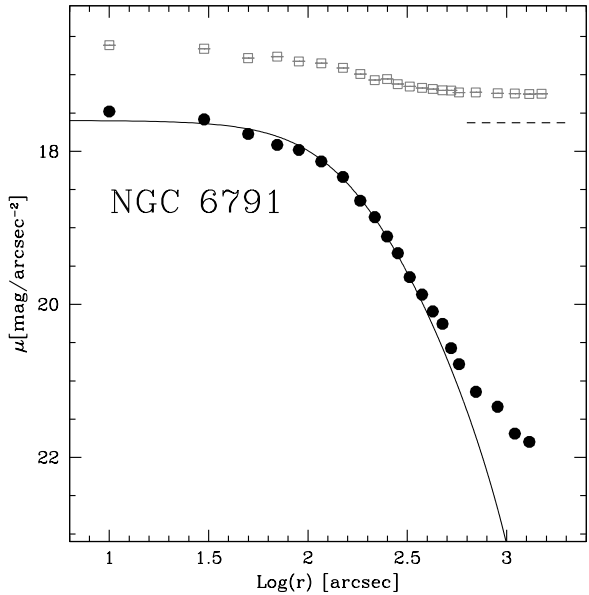}
\caption{Optical instrumental surface brightness profile for NGC~6791.The grey empty squares represent the
observed raw profile, while the black circles are the sky subtracted values. The horizontal dashed line
marks the estimated of the background level as inferred in the North-West quadrant. The background-subtracted 
profile is shown with filled black circles.}
\label{map}
\end{figure}

\section{Mass loss and initial mass estimate.}

The observational evidence presented in the previous Sections reveal that NGC~6791 
might have lost a fraction of its original mass due to environmental effects (like 
tidal shocks due to close interactions with giant molecular clouds, spiral arms, the Galactic disk 
and, in general, to the interactions with the Galactic tidal field) or 
internal dynamical effects (like two-body relaxation.
In this Section we attempt to estimate the total
mass lost by NGC~6791 during its evolution and derive its original mass.

While many recipes can be used for this purpose (e.g., Vesperini et al. 2013), we adopted 
the approach described by Lamers et al. (2005).
It has the advantage of describing the way the
mass of a cluster decreases with time by means of relatively simple analytic expressions.
The analysis by Lamers et al. (2005) accounts for the effect of both stellar and dynamical evolution.
It is important to note that the results obtained with this simplified approach are
in good agreement with those obtained by detailed $N$-body simulations for clusters in the tidal field of the
Galaxy (see also Lamers \& Gieles 2006).

Lamers et al. (2005) provide an expression for the initial mass of the cluster, 
$M_{\rm ini}$, in terms of its age ($t$) and present mass ($M$):

\vspace{0.5cm}

\begin{equation} 
\label{eq:m_ini_approx}
M_{\rm ini} \simeq \left[
    \left( \frac{M}{\mathrm{M}_{\odot}} \right)^{\gamma} + 
    \frac{\gamma t}{t_0} \right]^{1/\gamma} \left[1-q_{\mathrm{ev}}(t) \right]^{-1}
\end{equation}

\vspace{0.5cm}

\noindent $t_0$ is the dissolution time scale parameter. It is a constant depending on the strength of the tidal field
and it basically describes the mass loss 
due to dissolution processes such as Galactic tidal field interactions and shocks due to encounters with 
giant molecular clouds or spiral arms (Gieles et al. 2006, 2007).
The smaller values of $t_0$ are typically associated to encounters with molecular clouds and spiral arms, while
the larger to the Galactic tidal field. Also, the longer $t_0$, the weaker the tidal effect is. 
We used  $t_0=3.3_{-1.0}^{+1.4}$\,Myr 
(corresponding to the disruption time of $1.3\pm0.5$\,Gyr for 
a cluster with the initial mass of $10^4$M$_{\odot}$), obtained by Lamers et al. (2005) by comparing the
distribution of mass and age of OCs in the solar neighbourhood with theoretical predictions. 
This choice is reasonable since 
the current Galactocentric distance of NGC~6791 roughly corresponds to that of the Sun.

$\gamma$ is a dimensionless index, which depends on the cluster initial density distribution 
and has typical values ranging from about 0.6 to 0.8, which increases with the concentration,
as constrained from theoretical studies (Gieles et al. 2004) and observations 
(Boutloukos et al. 2003).
$\gamma$ is usually constrained by 
the King dimensionless potential $W_0$, which characterises 
the concentration of the King models (King 1966).
We adopted here $\gamma = 0.62$ corresponding to $W_0=5$ that is a typical value for open clusters
and is in agreement with what derived by 
the analysis of the observed star count density and surface brightness profiles ($W_0=5.3\pm1.0$; Section~4).
However, it is important to note that the star cluster density distributions change with time, 
therefore including the present-day value of $W_0$ in Eq.~(\ref{eq:m_ini_approx}) represents 
a crude approximation. 
In particular, it is well known (e.g. Trenti et al. 2010) that the concentration of stellar 
systems tend to increase (and $W_0$ accordingly increases) 
as the dynamical evolution of star cluster proceeds. As a consequence, given the proportionality 
linking $\gamma$ to $W_0$, if the adopted value of $W_0$ is larger that its initial one, also 
the adopted value of $\gamma$ will represent an upper limit to the correct one.

The function $q_{\mathrm{ev}}(t)$ describes the mass loss due to stellar evolution and can be 
approximated by the following analytical formula:

\begin{equation} 
\label{eq:mass_loss_ev}
\log_{10} q_{\mathrm{ev}}(t) = (\log_{10} t - a)^{b} + c, \quad {\rm for}\,\,t > 12.5\,\,\mathrm{Myr},
\end{equation}

\noindent where $a$, $b$, and $c$ are coefficients that depend on the cluster metallicity.
The [Fe$/$H] abundance of NGC\,6791 is about $+0.4$ ($Z\sim0.05$; Carraro et al. 2006, Gratton et al. 2006,
 Origlia et al. 2006) leading
(see Table~1 and Eq.\,2 by Lamers et al. 2005) to values of $a=7.00$, $b=0.25$, and $c=-1.82$.

\begin{figure}
  \centering
  \includegraphics[width=85mm]{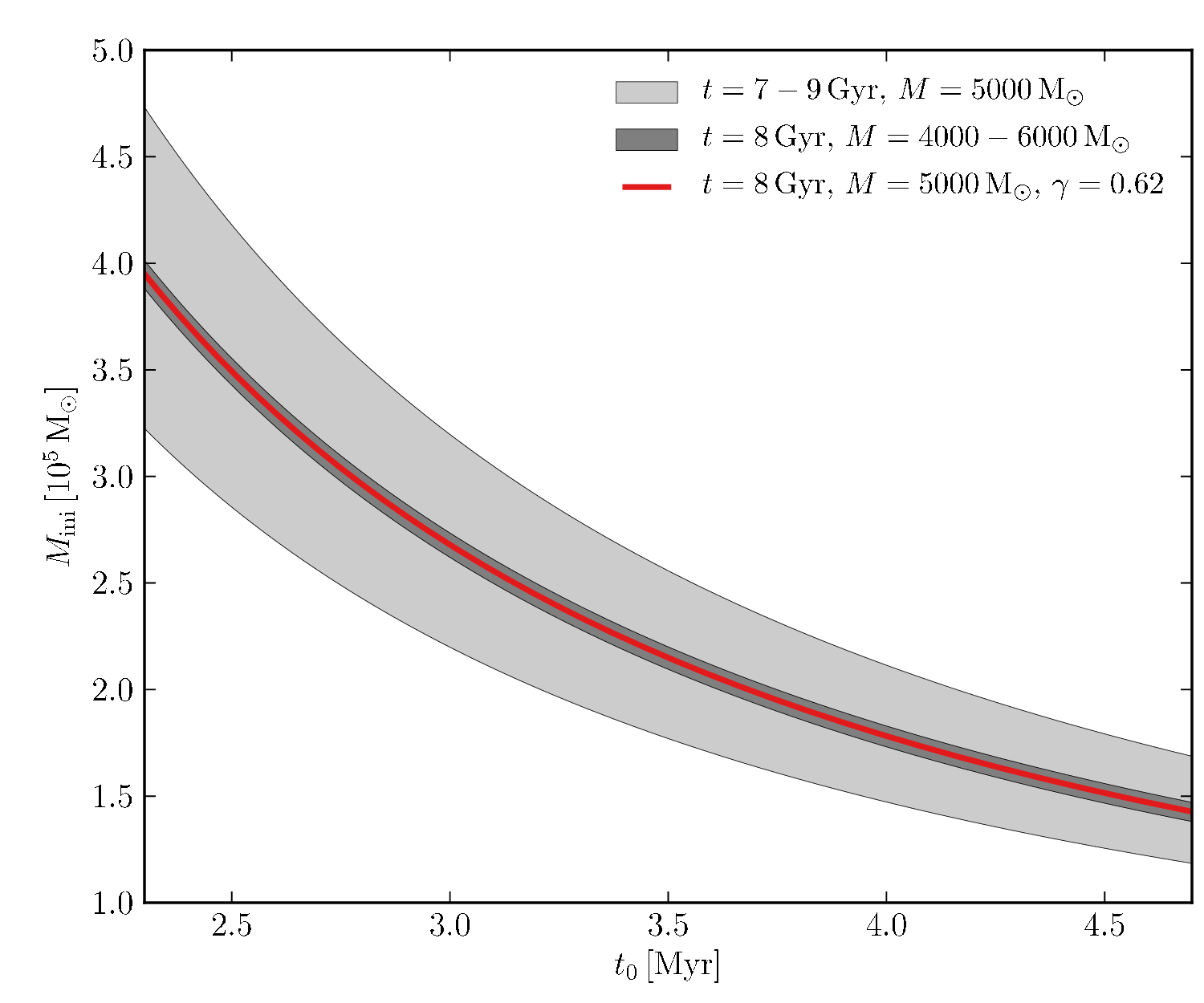}
  \caption{Cluster initial mass given by Eq.~(\ref{eq:m_ini_approx}) as a function of the dissolution time scale parameter $t_0$. 
  The red curve shows the dependence derived for the current cluster mass of 5000\,M$_{\odot}$ and its age
  of 8\,Gyr; the dark-grey coloured 
  area corresponds to values derived for mass within 4000--6000\,M$_{\odot}$ at age of 8\,Gyr; the light-grey area corresponds to values derived for 
  the mass of 5000\,M$_{\odot}$ at an age within 7--9\,Gyr.}
  \label{fig:mini_t0}
\end{figure}

Eq.~(\ref{eq:m_ini_approx}) depends also on the cluster current mass $M$ and its current age $t$. 
We adopted a current mass $M=5000\,\mathrm{M}_{\odot}$ as derived by Platais et al. (2011) and age $t=8$ Gyr 
(Grundahl et al. 2008, Garcia-Berro et al. 2010, Brogaard et al. 2012). 
These values posses significant errors and their exact estimate is still a matter of debate,
however it is important to note that the final error in the initial mass estimate 
is driven mostly by the uncertainty in the dissolution time scale parameter $t_0$. 
The dependence of the cluster initial mass on $t_0$ within the range 
of  $3.3_{-1.0}^{+1.4}$\,Myr is plotted in Fig.~\ref{fig:mini_t0}, together with the spread caused by the 
uncertainty of the current cluster mass (lower and upper limit of 4000 and 6000\,M$_{\odot}$ considered) 
and its current age (lower and upper limit of 7 and 9\,Gyr considered).

The initial mass of NGC\,6791 obtained by Eq.~(\ref{eq:m_ini_approx}) turns out to be  
$M_{\rm ini}=(1.5$--$4) \times 10^5\, M_{\odot}$, i.e. more than 50 times larger than its present-day mass.
For reasonable smaller values of $W_0$\footnote{Note that OCs with very small values of the dimensionless 
potential ($W_0<3$), are expected to be completely destroyed in a short timescale ($\sim 10^7$ yr).} 
the initial mass derived by means of Eq.~(\ref{eq:m_ini_approx}) 
will be slightly larger, but still compatible within the errors with the value quoted above. 

It is important to stress that this estimate is based on a simplified approach and on 
parameters derived by the average behaviours of OCs. 
Different recipes and assumptions may likely lead to slightly different results.

\section{Conclusions}

In this study we have presented a deep, wide field, photometric data-set extending well
beyond the tidal radius of the massive open cluster NGC~6791.  
This data-set allowed us to derive updated estimates of its structural parameters, 
by fitting a King profile over both the star counts and surface brightness profiles.
With respect to previous studies, that were limited in spatial coverage (Platais et al. 2011),
the King model best fitting the projected density profile turns out to have larger values of both c
and $r_t$.
 
More interestingly, for the first time, we detected clear signatures of tidal features in the star distribution of
NGC~6791, in the form 
of irregular, but evident, elongation and tidal tails. Some of them seem to closely follow the cluster motion,
while others are nearly perpendicular to the Galactic disk. 
These are clearly visible both in the 2D surface density map and in the projected density profile and they 
represent typical evidence of recent stellar mass loss.
We can therefore argue that NGC~6791 might have lost quite a significant fraction 
of its original mass while orbiting around the Galactic center. 
By using a simple analytic approach (Lamers et al. 2005), we estimated the mass likely lost by NGC~6791
during its evolution because of the effect of both stellar evolution and dynamical interactions. 
From this calculation we estimated the mass at its birth to be
$M_{\rm ini}=(1.5$--$4) \times 10^5\, M_{\odot}$, several tens larger than its present-day mass.

This derivation lends support to a scenario in which NGC 6791 was a much more 
massive system. 
This would qualitatively explain why the cluster could have survived for such a long time 
(its age is around 7-8 Gyr)
contrary to the expectations of current estimates for the destruction rates of OCs in the Galaxy.

It is interesting to note also that 
$M_{\rm ini}$ is comparable to the present-day mass of most Galactic globular clusters and it
is in rough agreement with the initial mass of the least massive clusters hosting multiple stellar 
populations  (see for example Dalessandro et al. 2014). 
Therefore, the estimate of $M_{\rm ini}$ obtained in this paper might set interesting constraints on 
the initial conditions of stellar systems able to form multiple stellar populations.

\vspace{1.0cm}
\noindent  {\bf ACKNOWLEDGMENTS}

\noindent  ED thanks Michele Bellazzini for useful discussions and suggestions.
The authors thank the anonymous referee for the careful reading of the paper and his/her suggestions.


\begin{thebibliography}{}

\bibitem[Baumgardt 
\& Makino(2003)]{baumgardt03} Baumgardt, H., \& Makino, J.\ 2003, MNRAS, 340, 227

\bibitem[Bedin et 
al.(2006)]{2006A&A...460L..27B} Bedin, L.~R., Piotto, G., Carraro, G., King, I.~R., \& Anderson, J.\ 2006, A\&A, 460, L27 


\bibitem[Bensby et 
al.(2013)]{2013A&A...549A.147B} Bensby, T., Yee, J.~C., Feltzing, S., et al.\ 2013, A\&A, 549, AA147 

\bibitem[Bergond et al.(2001)]{2001A&A...377..462B} Bergond, G., Leon, S., \& Guibert, J.\ 2001, A\&A, 377, 462   

\bibitem[Binney \& Tremaine(1987)]{1987gady.book.....B} Binney, J., \& Tremaine, S.\ 1987, Princeton, NJ, 
Princeton University Press, 1987, 747 p.,


\bibitem[Boutloukos 
\& Lamers(2003)]{2003MNRAS.338..717B} Boutloukos, S.~G., \& Lamers, H.~J.~G.~L.~M.\ 2003, MNRAS, 338, 717 

\bibitem[Bragaglia et al.(2014)]{2014ApJ...796...68B} Bragaglia, A., 
Sneden, C., Carretta, E., et al.\ 2014, ApJ, 796, 68 

\bibitem[Brogaard et 
al.(2012)]{2012A&A...543A.106B} Brogaard, K., VandenBerg, D.~A., Bruntt, H., et al.\ 2012, A\&A, 543, AA106 


\bibitem[Buzzoni et al.(2012)]{2012ApJ...749...35B} Buzzoni, A., Bertone, 
E., Carraro, G., \& Buson, L.\ 2012, ApJ, 749, 35

\bibitem[Capuzzo Dolcetta et al.(2005)]{2005AJ....129.1906C} Capuzzo 
Dolcetta, R., Di Matteo, P., \& Miocchi, P.\ 2005, AJ, 129, 1906 

\bibitem[Carraro et al.(2006)]{2006ApJ...643.1151C} Carraro, G., Villanova, 
S., Demarque, P., et al.\ 2006, ApJ, 643, 1151

\bibitem[Carraro (2014)]{2014ASPC..482..245C} Carraro, G., \ 2014, ASP Conference Series, Vol. 482., p. 245


\bibitem[Combes et 
al.(1999)]{1999A&A...352..149C} Combes, F., Leon, S., \& Meylan, G.\ 1999, A\&A, 352, 149 


\bibitem[Cunha et al.(2015)]{2015ApJ...798L..41C} Cunha, K., Smith, V.~V., 
Johnson, J.~A., et al.\ 2015, ApJL, 798, LL41 


\bibitem[Dalessandro et al.(2009)]{2009ApJS..182..509D} Dalessandro, E., 
Beccari, G., Lanzoni, B., et al.\ 2009, ApJS, 182, 509

\bibitem[\protect\citeauthoryear{Dalessandro et 
al.}{2012}]{2012AJ....144..126D} Dalessandro E., Schiavon R.~P., Rood 
R.~T., Ferraro F.~R., Sohn S.~T., Lanzoni B., O'Connell R.~W., 2012, AJ, 
144, 126 

\bibitem[Dalessandro et al.(2013)]{2013ApJ...778..135D} Dalessandro, E., 
Ferraro, F.~R., Massari, D., et al.\ 2013, ApJ, 778, 135

\bibitem[Dalessandro et al.(2014)]{2014ApJ...791L...4D} Dalessandro, E., 
Massari, D., Bellazzini, M., et al.\ 2014, ApJL, 791, LL4 

\bibitem[Davenport 
\& Sandquist(2010)]{2010ApJ...711..559D} Davenport, J.~R.~A., \& Sandquist, E.~L.\ 2010, ApJ, 711, 559 


\bibitem[Garc{\'{\i}}a-Berro et al.(2010)]{2010Natur.465..194G} 
Garc{\'{\i}}a-Berro, E., Torres, S., Althaus, L.~G., et al.\ 2010, Nature, 
465, 194 


\bibitem[Geisler et al.(2012)]{2012ApJ...756L..40G} Geisler, D., Villanova, 
S., Carraro, G., et al.\ 2012, ApJL, 756, LL40 

\bibitem[Gieles et al.(2007)]{gieles07} Gieles, M., 
Athanassoula, E., \& Portegies Zwart, S.~F.\ 2007, MNRAS, 376, 809 


\bibitem[Gieles et al.(2006)]{gieles06} Gieles, M., Portegies 
Zwart, S.~F., Baumgardt, H., et al.\ 2006, MNRAS, 371, 793 


\bibitem[Gieles et al.(2004)]{gieles04} Gieles, M., Baumgardt, 
H., Bastian, N., 
\& Lamers, H.~J.~G.~L.~M.\ 2004, The Formation and Evolution of Massive Young Star Clusters, 322, 481 

\bibitem[Gratton et al.(2006)]{2006ApJ...642..462G} Gratton, R., Bragaglia, 
A., Carretta, E., \& Tosi, M.\ 2006, ApJ, 642, 462 



\bibitem[Grundahl et 
al.(2008)]{2008A&A...492..171G} Grundahl, F., Clausen, J.~V., Hardis, S., \& Frandsen, S.\ 2008, A\&A, 492, 171 


\bibitem[J{\'{\i}}lkov{\'a} et 
al.(2012)]{jil12} J{\'{\i}}lkov{\'a}, L., Carraro, G., Jungwiert, B., \& Minchev, I.\ 2012, A\&A, 541, AA64 


\bibitem[Johnston et al.(1999)]{1999ApJ...512L.109J} Johnston, K.~V., Zhao, 
H., Spergel, D.~N., \& Hernquist, L.\ 1999, ApJL, 512, L109 


\bibitem[King(1966)]{1966AJ.....71...64K} King, I.~R.\ 1966, AJ, 71, 64 

\bibitem[King et al.(2005)]{2005AJ....130..626K} King, I.~R., Bedin, L.~R., 
Piotto, G., Cassisi, S., \& Anderson, J.\ 2005, AJ, 130, 626

\bibitem[Kinman(1965)]{1965ApJ...142..655K} Kinman, T.~D. \ 1965, ApJ, 142, 655

\bibitem[Kuhn et al.(1996)]{1996ApJ...469L..93K} Kuhn, J.~R., Smith, H.~A., 
\& Hawley, S.~L.\ 1996, ApJL, 469, L93 


\bibitem[K{\"u}pper et al.(2010)]{2010MNRAS.407.2241K} K{\"u}pper, 
A.~H.~W., Kroupa, P., Baumgardt, H., 
\& Heggie, D.~C.\ 2010, MNRAS, 407, 2241

\bibitem[Lamers et 
al.(2005)]{lamers05} Lamers, H.~J.~G.~L.~M., Gieles, M., Bastian, N., et al.\ 2005, A\&A, 441, 117 

\bibitem[Lamers \& Gieles (2006)]{lamers06} Lamers, H.~J.~G.~L.~M., Gieles, M.\ 2006, A\&A, 455, 17 

\bibitem[Leon(1998)]{1998PhDT.......217L} Leon, S.\ 1998, Ph.D.~Thesis,  


\bibitem[Liebert et al.(1994)]{1994AJ....107.1408L} Liebert, J., Saffer, 
R.~A., \& Green, E.~M.\ 1994, AJ, 107, 1408 


\bibitem[Miocchi et al.(2013)]{2013ApJ...774..151M} Miocchi, P., Lanzoni, 
B., Ferraro, F.~R., et al.\ 2013, ApJ, 774, 151 

\bibitem[Montuori et al.(2007)]{2007ApJ...659.1212M} Montuori, M., 
Capuzzo-Dolcetta, R., Di Matteo, P., Lepinette, A., 
\& Miocchi, P.\ 2007, ApJ, 659, 1212 

\bibitem[Odenkirchen et al.(2001)]{2001ApJ...548L.165O} Odenkirchen, M., 
Grebel, E.~K., Rockosi, C.~M., et al.\ 2001, ApJL, 548, L165 


\bibitem[Odenkirchen et al.(2003)]{2003AJ....126.2385O} Odenkirchen, M., 
Grebel, E.~K., Dehnen, W., et al.\ 2003, AJ, 126, 2385 

\bibitem[Origlia et al.(2006)]{2006ApJ...646..499O} Origlia, L., Valenti, 
E., Rich, R.~M., \& Ferraro, F.~R.\ 2006, ApJ, 646, 499 


\bibitem[Platais et al.(2011)]{2011ApJ...733L...1P} Platais, I., Cudworth, 
K.~M., Kozhurina-Platais, V., et al.\ 2011, ApJL, 733, LL1


\bibitem[Rockosi et al.(2002)]{2002AJ....124..349R} Rockosi, C.~M., 
Odenkirchen, M., Grebel, E.~K., et al.\ 2002, AJ, 124, 349 


\bibitem[Sollima et al.(2011)]{2011ApJ...726...47S} Sollima, A., Mart{\'{\i}}nez-Delgado, D., Valls-Gabaud, D., \& Pe{\~n}arrubia, J.\ 2011, ApJ, 726, 47 

\bibitem[Spitzer & Harm(1958)]{1958ApJ...127..544S} Spitzer, L., Jr., \& Harm, R.\ 1958, ApJ, 127, 544 


\bibitem[Spitzer(1958)]{1958ApJ...127...17S} Spitzer, L., Jr.\ 1958, ApJ, 127, 17 


\bibitem[Stetson(1987)]{1987PASP...99..191S} Stetson, P.~B.\ 1987, PASP, 
99, 191 

\bibitem[Tofflemire et al.(2014)]{2014AJ....148...61T} Tofflemire, B.~M., 
Gosnell, N.~M., Mathieu, R.~D., \& Platais, I.\ 2014, AJ, 148, 61 

\bibitem[Trenti et al.(2010)]{2010ApJ...708.1598T} Trenti, M., Vesperini, 
E., \& Pasquato, M.\ 2010, ApJ, 708, 1598 

\bibitem[Twarog et al. (2011)]{2011ApJ...727L...7T} Twarog, B.A., Carraro, G., Anthony-Twarog, B.J, \ 2011,
ApJ, 727, L7

\bibitem[Vesperini et al.(2013)]{2013MNRAS.429.1913V} Vesperini, E., 
McMillan, S.~L.~W., D'Antona, F., \& D'Ercole, A.\ 2013, MNRAS, 429, 1913 


\bibitem[Wilson(1975)]{1975AJ.....80..175W} Wilson, C.~P.\ 1975, AJ, 80, 
175 

\end{thebibliography}
\end{document}